\documentclass[a4paper,fleqn]{cas-dc}

\usepackage[numbers]{natbib}
\usepackage[switch]{lineno}
\newcommand{\absdiv}[1]{%
  \par\addvspace{.5\baselineskip}
  \noindent\textit{#1}\quad\ignorespaces
}
\usepackage{tabularx}
\usepackage{amsmath}
\usepackage{soul,color}
\soulregister\cite7
\usepackage{graphicx}
\usepackage{dcolumn}
\usepackage{array}
\usepackage{algorithm} 
\usepackage{algpseudocode} 
\usepackage{mathtools}
\usepackage{siunitx}
\newcolumntype{L}[1]{>{\raggedright\arraybackslash}p{#1}}
\newcolumntype{C}[1]{>{\centering\arraybackslash}p{#1}}
\newcolumntype{R}[1]{>{\raggedleft\arraybackslash}p{#1}}

\begin{document}

\let\WriteBookmarks\relax
\def\floatpagepagefraction{1}
\def\textpagefraction{.001}
\shorttitle{Characterization of wetting using topological principles}
\shortauthors{C. Sun et~al.}

\title [mode = title]{Characterization of wetting using topological principles}                   

\author[1]{Chenhao Sun}
\author[2]{James E. McClure}
\author[1]{Peyman Mostaghimi}
\author[3]{Anna L. Herring}
\author[4]{Douglas E. Meisenheimer}
\author[4]{Dorthe Wildenschild}
\author[5,6,7]{Steffen Berg}
\author[1]{Ryan T. Armstrong}
\cormark[1]
\ead{E-mail address: ryan.armstrong@unsw.edu.au}
\address[1]{\normalsize \itshape School of Minerals \& Energy Resources Engineering, University of New South Wales, Kensington, NSW 2052, Australia}
\address[2]{\itshape Advanced Research Computing, Virginia Polytechnic Institute \& State University, Blacksburg, Virginia 24061, USA}
\address[3]{\itshape Department of Applied Mathematics, Australian National University, Canberra, ACT 2601, Australia}
\address[4]{\itshape Department of Chemical, Biological and Environmental Engineering, Oregon State University, Corvallis, Oregon 97331, USA}
\address[5]{\itshape Hydrocarbon Recovery, Shell Global Solutions International B.V., Grasweg 31, 1031 HW Amsterdam, The Netherlands}
\address[6]{\itshape Department of Earth Science \& Engineering, Imperial College London, London SW7 2AZ, UK}
\address[7]{\itshape Department of Chemical Engineering, Imperial College London, London SW7 2AZ, UK}
\cortext[cor1]{Corresponding author}

\begin{abstract}
\absdiv{\it Hypothesis}
\newline
Understanding wetting behavior is of great importance for natural systems and technological applications. The traditional concept of contact angle, a purely geometrical measure related to curvature, is often used for characterizing the wetting state of a system. It can be determined from Young's equation by applying equilibrium thermodynamics. However, whether contact angle is a representative measure of wetting for systems with significant complexity is unclear. Herein, we hypothesize that topological principles based on the Gauss-Bonnet theorem could yield a robust measure to characterize wetting.
\absdiv{\it Theory and Experiments}
\newline
We introduce a macroscopic contact angle based on the deficit curvature of the fluid interfaces that are imposed by contacts with other immiscible phases. We perform sessile droplet simulations followed by multiphase experiments for porous sintered glass and Bentheimer sandstone to assess the sensitivity and robustness of the topological approach and compare the results to other traditional approaches.  
\absdiv{\it Findings}
\newline
We show that the presented topological principle is consistent with thermodynamics under the simplest conditions through a variational analysis. Furthermore, we elucidate that at sufficiently high image resolution the proposed topological approach and local contact angle measurements are comparable. While at lower resolutions, the proposed approach provides more accurate results being robust to resolution-based effects. Overall, the presented concepts open new pathways to characterize the wetting state of complex systems and theoretical developments to study multiphase systems. 
\end{abstract}


\begin{keywords}
Wetting behavior \sep Geometric state of fluids \sep Topological principles \sep Gaussian curvature \sep Porous media \sep Multiphase flow \sep Interfacial curvature 
\end{keywords}

\maketitle
\section{Introduction}

The inherent property of wetting refers to the preferential affinity of a fluid that is immersed in another immiscible fluid to coat a solid material \cite{de1985wetting,bonn2009wetting,de2013capillarity}. Due to the ubiquitous existence of colloidal and interfacial phenomena in nature and applications, understanding the role of wetting is a key principal of interest. This intriguing interest in wetting behavior is motivated by numerous advanced technologies in nanotechnology, biological engineering, material science and geosciences \cite{powell2011electric,xu2014proteins,blossey2003self,bartels2017oil}. For instance, the design of bio-inspired fluidics, directional fluid transportation, composite functional materials, nano/micro-fluidics and energy storage systems require a precise and effective way to describe the wetting state therein. 

\begin{figure}
\hspace*{-0.2cm}\includegraphics[width=0.51\textwidth]{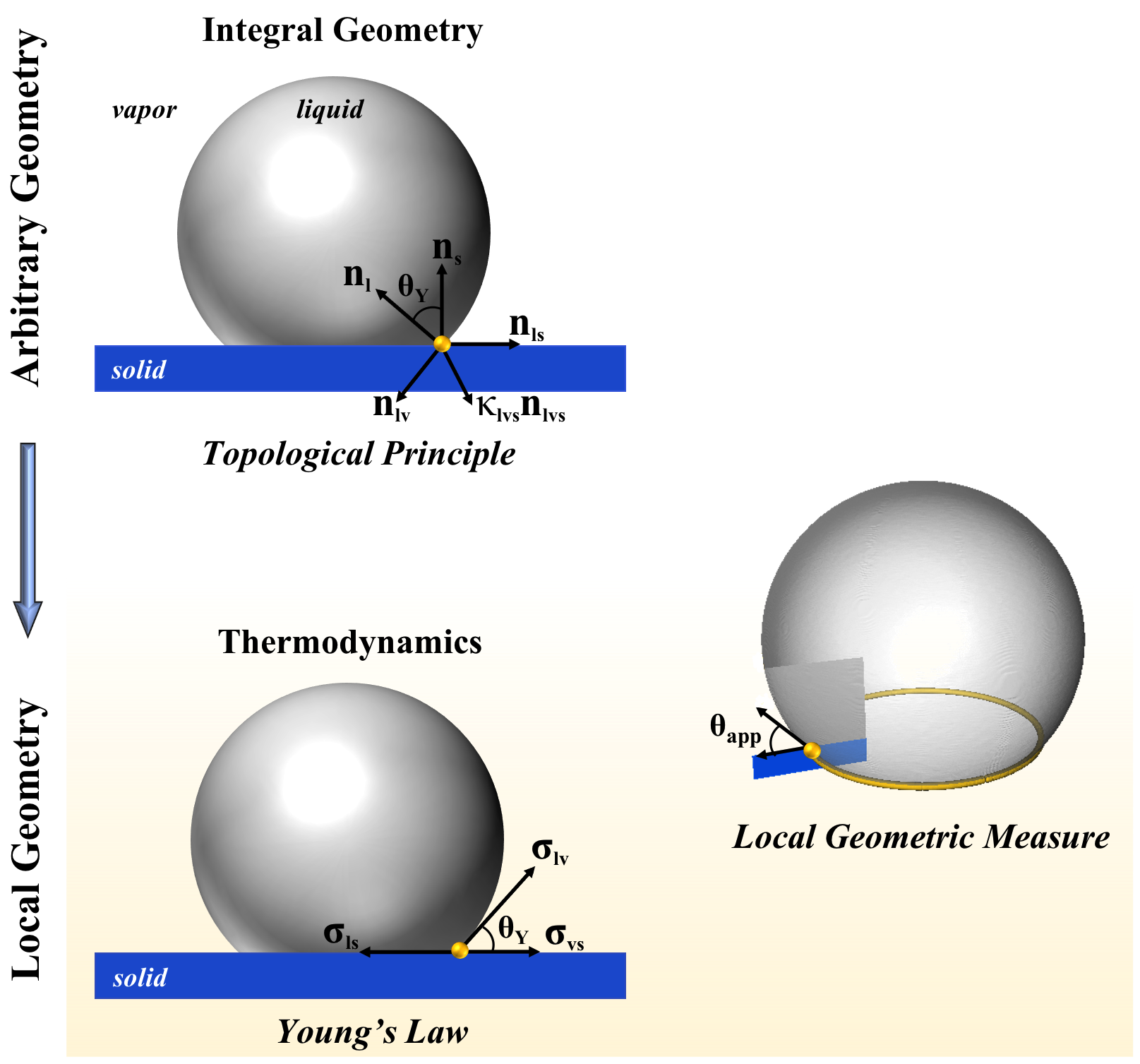}
\caption{The schematic illustration of topological principle based on integral geometry \cite{sun2020probing} and Young's equation based on thermodynamics at a microscopic contact point when a sessile droplet is deposited on a flat solid substrate exposed to vapor phase. The schematic illustration of apparent contact angle $\theta_{app}$ extraction by local measurement along the contact line.}
\label{fig1}
\end{figure}

Since 1805, Young's equation, based on thermodynamic laws, is firmly established to infer the wettability of a flat and chemically homogeneous solid surface at equilibrium \cite{young1805iii}, $\cos \theta_Y  = \frac{\sigma_{ls} - \sigma_{vs}}{\sigma_{lv}}$, where the subscripts denote the immiscible fluids (liquid $l$ and vapor $v$) and solid ($s$) for the associated surface free energies. $\theta_Y$ is the contact angle formed at the microscopic contact point on the surface, as depicted in Fig. \ref{fig1}. Wetting hysteresis due to the contribution of surface heterogeneity and contact line dynamics has been studied in detail over the last century \cite{wenzel1936resistance,cassie1944wettability,decker1997contact}. However, despite theoretical and experimental investigations through the last 70 years \cite{de1985wetting,bonn2009wetting,yu2015wetting,turmine2000thermodynamic,blunt2019thermodynamically,mahani2015kinetics}, several fundamental challenges in characterizing wetting state of complex multiphase systems remain unsolved and are currently pending. 

For disordered and complex porous geometries ranging from catalysts in fuel cells to lungs in the respiratory system to porous glass filters to subsurface rocks, there are multiple length scales, various surface free energies, and surface heterogeneity involved. The convoluted interplay of physicochemical properties with multi-scale complexities therein presents significant spatial variability of contact angles along the three-phase contact line \cite{sun2020probing,holtzman2015wettability}, which leads to the resulting wetting hysteresis behavior and pinning effects. Accordingly, these features trigger a formidable challenge to characterizing wetting behavior, which are far from being solved. It is therefore controversial whether Young's law is still applicable for these disordered and complex solids \cite{garfi2019fluid}. 

In the past, contact angle (and curvature) measurements were mainly based on two-dimensional (2D) projections, and in configurations that are optically transparent such as a typical sessile drop setup \cite{kwok1997contact}. However, three-dimensional (3D) imaging provides the possibility to determine contact angles within opaque porous media \cite{scanziani2017automatic,alratrout2017automatic,ibekwe2020automated,dalton2018methods}. Possible 3D imaging technologies are 3D X-ray microcomputed tomography (micro-CT) \cite{wildenschild2013x} and confocal microscopy \cite{sundberg2007contact}. Given the rapid developments in 3D imaging technologies, there is an opportunity for interface science to make use of this methodology where the wetting of various porous domains would be of interest. The first community that started adopting this new technology is the porous media community, out of pure necessity to measure the wettability of immiscible fluids in geological materials, which are naturally opaque \cite{andrew2014pore,alratrout2018wettability,tudek2017situ}. These efforts rely on micro-CT images of porous rocks saturated with immiscible phases \cite{wildenschild2013x}. The image voxels of a 3D system are segmented into respective phases \cite{schluter2014image} followed by the identification of the three-phase contact line to measure spatially distributed apparent contact angles, $\theta_{app}$, for each microscopic three-phase contact point along the contact line. As shown in the right side of Fig. \ref{fig1}, the method measures the angle between the local tangential plane of liquid/vapor interface and the solid surface in the vicinity of three-phase contact points. Thus, the approach represents \textit{in situ} $\theta_{app}$ directly along the contact line, which refers to microscopic wetting. However, the simplicity of $\theta_{app}$ computed by the local method is susceptible to errors due to image quality and pixelation-related segmentation errors \cite{armstrong2012linking,klise2016automated}. In particular, the identification of the three-phase contact line and measurement of an angle over a few voxels is inherently error-prone, as will be investigated herein. Ultimately, these systematic errors complicate the process of quantifying the wetting state of a porous multiphase system. 

Another issue arises by using for the purpose of wetting characterization the concept of contact angle, which is subject to hysteresis. Due to hysteresis that is emphasized by complex geometries, surface heterogeneity, and interface dynamics, it is very questionable whether contact angle is a representative measure for characterizing wetting in such systems \cite{wenzel1936resistance,cassie1944wettability,johnson1964contact,morrow1975effects,priest2007asymmetric}. In addition, the line tension and disjoining/cojoining pressure occur along the contact line. Contact line tension, which is the excess free energy per unit length, is a dominant parameter in microscopic wetting \cite{indekeu1994line}. It contributes to intermolecular force balance, resulting in contact line pinning and the associated hysteresis loop of contact angles \cite{de1985wetting,bonn2009wetting,de2013capillarity}. Therefore, the variations on contact line curvature and the associated $\theta_{app}$ vary from a microscopic point of view due to the effects of surface heterogeneity and flow dynamics, which yields unexpected wetting behavior. Despite these parameters influencing microscopic wetting and hysteresis behavior, the question arises as to whether $\theta_{app}$ can capture enough information to represent the wetting state of the system and provide sufficient guidance for the design of functional surfaces?

Previous studies demonstrate that $\theta_{app}$ alone provides an incomplete description of wetting where the contact line is asymmetric \cite{rabbani2018pore}; especially for characterizing the macroscopic wetting behavior of multiphase systems, which has remained unexplored from a theoretical perspective. The wetting behavior synergistic affects on the phase topology and contact area with the solid surface, which must also be considered. It can be representative of the macroscopic wetting of the system and captures the complete microscopic information related to the thermodynamics. To this end, in our previous works \cite{sun2020linking,sun2020probing}, we developed a theory based on topological principles that can effectively describe wetting behavior, which links across various length scales pertinent to wetting phenomena. Herein, we aim to quantitatively and qualitatively unravel the effective and robust nature of the proposed theory by providing in-depth analysis and comparison to other recent methods for characterizing the wettability of multiphase systems.  

\section{Theoretical Concepts}
\subsection{Deficit Curvature from Gaussian Curvature}
Topological principle is applied to characterize the wetting state of a given multiphase system by the link of the Gauss-Bonnet theorem. For a fluid droplet $D$ that has a closed interface in the three-phase system, where liquid ($l$), vapor ($v$) and solid ($s$) are present, the total curvature of the fluid surface and its global topology are related by the Gauss-Bonnet theorem \cite{chern1944simple,sun2020linking}. As a consequence, the Euler characteristic $\chi$ and its total curvature for the surface $I$ of the droplet $D$ obey the following expression,

\begin{equation} \label{eq:1}
 2\pi \chi(D)=\int_{I} \kappa_G dA + \int_{\partial I} \kappa_g dC,
\end{equation}
where $dA$ and $dC$ are the droplet interfacial area element along the surface $I$ and the line element along the contact line $\partial I$, respectively. $\kappa_G$ and $\kappa_g$ are Gaussian curvature along the droplet interface and geodesic curvature along the contact line $\partial I$, respectively. 
\begin{figure*}
\centering\includegraphics[width=0.85\textwidth]{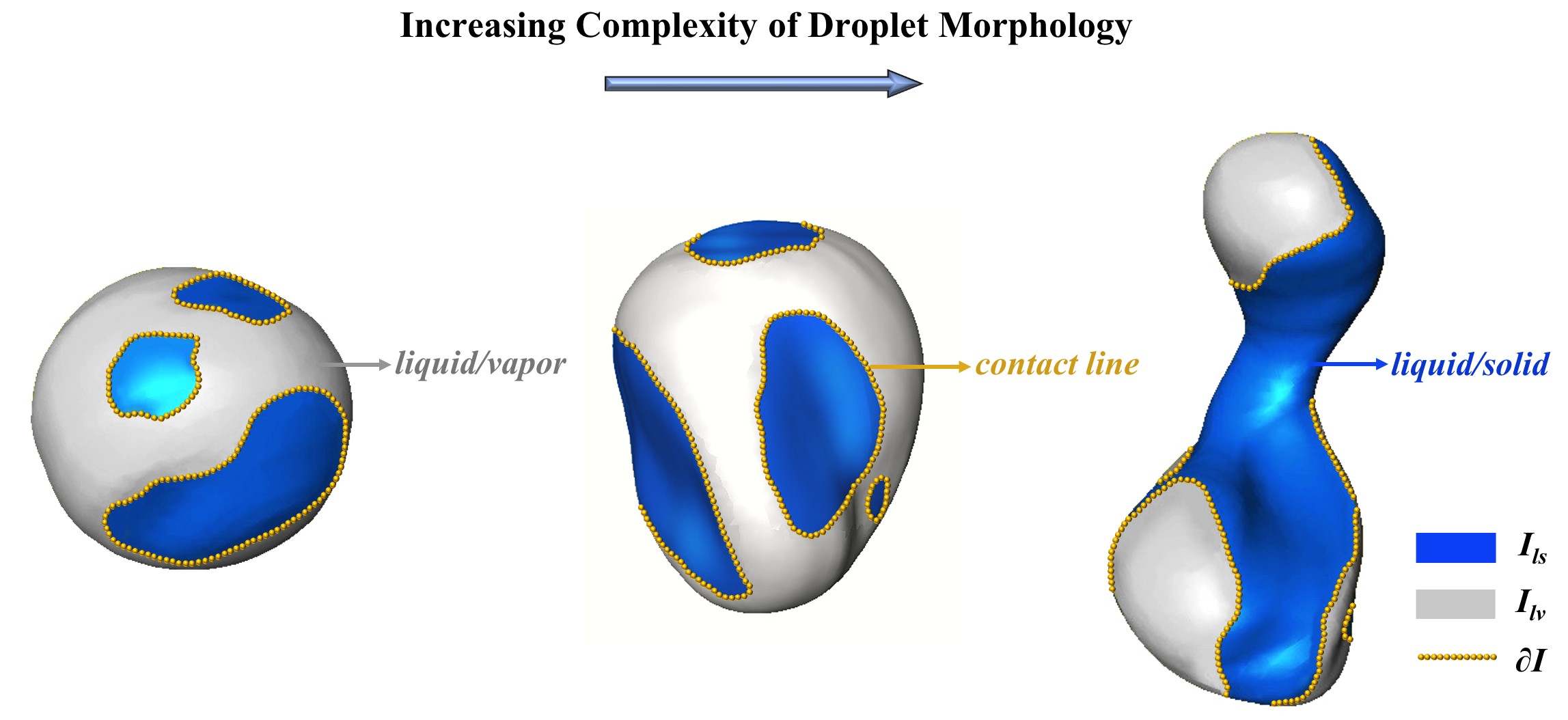}
\caption{Schematic diagram of 3D droplets on contact with a complex solid (transparent) with labeled definitions where grey color denotes the liquid/vapor interface ($I_{lv}$), and blue color denotes the liquid/solid interface ($I_{ls}$). The contact line loops are formed by three-phase contact points (yellow circles). From left to right, there is an increasing complexity of droplet morphology.}
\label{fig2}
\end{figure*}

For a 3D droplet, we can arrive at a generalized form of the expression by subdividing the fluid surface into liquid/vapor ($lv$) and liquid/solid ($ls$) interfaces,

\begin{equation} \label{eq:2}
\begin{split}
 4\pi\chi(D) &=2 \pi \chi(I_{lv}) + 2 \pi \chi (I_{ls}) \\
 &=\int_{I_{lv}} \kappa_{G}dA+\int_{I_{ls}} \kappa_{G}dA +\int_{\partial I}(\kappa_{g_{lv}}+\kappa_{g_{ls}})dC. 
\end{split}
\end{equation}

\noindent It indicates that the Euler characteristic of the droplet always remains constant and the geodesic curvature source term, $\int_{\partial I}(\kappa_{g_{lv}}+\kappa_{g_{ls}})dC$, will change accordingly based on the contribution of total surface curvature due to the wetting behavior of the system. The deficit curvature, $\Theta$, is defined as the summation of geodesic curvatures along the contact line relative to the tangential plane for each interface and corresponds to a total angle of change. By considering a droplet deposited on a flat, smooth and homogeneous surface as shown in the top of Fig.\ref{fig1}, $\Theta$ along the contact line can be expressed by the labeled notations as,

\begin{equation} \label{eq:3}
\begin{split}
\Theta &= \int_{\partial I}(\kappa_{g_{lv}}+\kappa_{g_{ls}})dC \\
&=\int_{\partial I} \kappa_{lvs}\mathbf{n}_{lvs}\cdot \big[\mathbf{n}_s \sin \theta_Y + \mathbf{n}_{ls} (1-\cos \theta_Y) \big]dC,
\end{split}
\end{equation}

\noindent where $\mathbf{n}_{lvs}$ is the normal vector to the contact line, which points in the direction of the curvature for the contact line. $\mathbf{n}_{lv}$ is the outward normal vector along the contact line relative to the droplet interface. $\mathbf{n}_{ls}$ is the outward normal vector along the contact line relative to the liquid/solid interface. From Eq. (\ref{eq:3}), it is evident that $\Theta$ obtained by applying the topological principle is an explicit average of intrinsic contact angle, i.e. Young's angle ($\theta_Y$) for this situation. Therefore, $\Theta$ can be interpreted in terms of fluid morphology in a way that is not affected by the contact angle hysteresis due to surface heterogeneity, geometry complexity, and interface dynamics \cite{sun2020probing}.

For solid surfaces that contain disordered and complex geometries or even with confined domains as shown in Fig. \ref{fig2}, we can revisit Eq. (\ref{eq:2}) to obtain $\Theta$. Macroscopic contact angle $\theta_{macro}$ can be obtained by using a normalizing factor $\lambda$ to scale the desired contact angle interval in $\theta_{macro}$ $\in [0,\pi]$,

\begin{equation} \label{eq:4}
\begin{split}
\theta_{macro} &= \lambda \Theta\\
&= \lambda \Big[4\pi\chi(D) - \Big(\int_{I_{lv}} \kappa_{G}dA+\int_{I_{ls}} \kappa_{G}dA \Big)\Big].
\end{split}
\end{equation}
The nature of this formulation resolves a few pressing issues. Firstly, the angle along the contact line is not directly computed as an average from the sequence of local geometric contact angle point measurements. $\theta_{macro}$ is inferred from interfacial curvature and area measurements along with a topological measurement. While these measures are still susceptible to pixelization effects, it is expected that these effects would be less than that resulting from local measurements along the three-phase contact line (as tested herein). Secondly, $\theta_{macro}$ accounts for the complete geodesic curvature of the contact line, which is the curvature of the three-phase contact line relative to both the solid surface and liquid/vapor interface. Therefore, the formulation captures wetting effects as evident by line tension and $\theta_{app}$, which are both known to influence wetting behavior. 

\subsection{Derivation of Young's Equation}
In this section, we show the direct link between the topological and thermodynamic concepts for determining the wetting state of the system by applying surface energy minimization and variational principles \cite{seo2015re}. The Gauss-Bonnet theorem allows us to express the total curvature of a droplet $D$ in terms of deficit curvature, corresponding average curvature and surface area, 

\begin{align}
4 \pi \chi(D) = \Theta + \kappa_{lv} A_{lv} + \kappa_{ls} A_{ls} \;.
\label{eq:5}
\end{align}

Here, we consider a variational principle of the internal energy as applied in thermodynamic approaches. The variation of the internal energy is given by Euler's homogeneous function theorem,

\begin{align}
\delta U = T \delta S - p_v \delta V_v - p_l\delta V_l + \sigma_{lv} \delta A_{lv}
\nonumber\\ + \sigma_{ls}\delta A_{ls} + \sigma_{vs} \delta A_{vs} \;.
\label{eq:6}
\end{align}
Based on this, we consider a closed system with $\delta U=0$. Droplet rearrangements can occur provided that the volume is not changed, meaning that $\delta V_l = \delta V_v = 0$. Therefore, entropy production in the system is strictly linked to the minimization of the surface energy,

\begin{align}
\delta S = - \frac 1 T \Big[ \sigma_{lv} \delta A_{lv}
+ \sigma_{ls}\delta A_{ls} + \sigma_{vs} \delta A_{vs} \Big] \ge 0 \;.
\label{eq:7}
\end{align}
The total surface area of the solid substrate is constant, then   

\begin{align}
\delta A_s = \delta A_{vs} + \delta A_{ls} = 0 \;,
\label{eq:8}
\end{align}
which can be used to eliminate one of the surface areas from Eq. (\ref{eq:7}). The topological constraint from Eq. (\ref{eq:5}) determines the condition that should be imposed to ensure that geometric variation occurs without changing the topology,

\begin{equation} \label{eq:9}
\begin{split}
\delta \big[ 4 \pi \chi(D)\big] & =  \delta \Theta + 
\kappa_{lv} \delta A_{lv} + A_{lv} \delta \kappa_{lv} \\ &\quad
\kappa_{ls} \delta A_{ls} + A_{ls} \delta \kappa_{ls} \\ & = 0
 \;.
\end{split}
\end{equation}

We shall impose Eq. (\ref{eq:9}) as a constraint on Eq. (\ref{eq:7}) using the method of Lagrange multipliers, also using Eq. (\ref{eq:8}) to eliminate $A_{vs}$. Now, we can express the entropy production as

\begin{equation}
\begin{split}
\delta S &= - \frac 1 T \Big[
\underbrace{\Big( \sigma_{ls} - \sigma_{vs}  + \frac{\sigma_{lv}\kappa_{ls}}{\kappa_{lv}} \Big)
}_{\mbox{surface area variation}}\delta A_{ls} \\
 &\quad + \frac{\sigma_{lv}}{\kappa_{lv}} 
\underbrace{\big( \delta \Theta + 
 A_{lv} \delta \kappa_{lv}  + A_{ls} \delta \kappa_{ls} \big)}_{\mbox{total curvature variation}} \Big]
 \;. 
\end{split}
 \label{eq:10}
\end{equation}
This expression separates terms associated with the variation of the surface area from terms that redistribute curvature along the cluster boundary. Put another way, the second term corresponds to a redistribution of the total curvature that occurs at constant surface area due to the deformation of the droplet. One of the ways to redistribute the total curvature is by altering the deficit curvature, which changes the contact angle. Since $T > 0$, the inequality can be re-expressed in the form,

\begin{equation}
\Big( \frac{\sigma_{ls} - \sigma_{vs}}{\sigma_{lv}} \kappa_{lv}  + \kappa_{ls} \Big) \delta A_{ls} 
+ \delta \Theta +  A_{lv} \delta \kappa_{lv} + A_{ls} \delta \kappa_{ls}  \le 0\;,
 \ 
\label{eq:11}
\end{equation}
which is equivalent to stating that the surface energy of the system must decrease.

\noindent Since the solid surface is flat, which means $\kappa_{ls} = 0$ and $\delta \kappa_{ls} = 0$. The variations can be computed directly based on expressions for a spherical cap with droplet radius $R$ and droplet height $h$ to the solid surface, where

\begin{equation}
\begin{split}
\Theta &= 2 \pi (1-\cos \theta)  \\
\kappa_{lv} &= \frac{1}{R^2} \\
A_{lv} &= 2 \pi R h \\
A_{ls} &= \pi h (2R-h)
\end{split}
\label{eq:12} 
\end{equation}

and the associated variations

\begin{equation} \label{eq:13}
\begin{split}
\delta \Theta &= \delta \big [ 2 \pi (1-\cos \theta) \big ] 
             = 2 \pi (h R^{-2}\delta R - R^{-1} \delta h )  \\
        \delta \kappa_{lv} &= \delta \big [ R^{-2} \big] = -2 R^{-3} \delta R \\
\delta A_{ls} &= \delta \big [\pi h (2R-h) \big] =
                    \pi \big( 2 h \delta R + 2 R \delta h - 2 h \delta h \big) 
\end{split}
\end{equation}
The volume of a spherical cap is $V = \frac 1 3 \pi h^2 \big( 3 R-h \big)$, and setting $\delta V=0$ imposes the relationship $\delta R = \Big( 1 - \frac{2R}{h} \Big) \delta h$. Inserting this into expressions above, we obtain

\begin{equation}
\begin{split}
\delta \Theta &= 2 \pi \Big(\frac{h}{R^2} - \frac{3}{R} \Big) \delta h 
\\
\delta \kappa_{lv} &= \Big(\frac{4}{R^2h} - \frac{2}{R^3} \Big) \delta h \\
\delta A_{ls} &=  -2 \pi R \delta h
\end{split}
\label{eq:14}
\end{equation}
Inserting Eqs. (\ref{eq:12}) and (\ref{eq:14}) into Eq. (\ref{eq:11}) and rearranging terms gives

\begin{eqnarray}
 \frac{2\pi \delta h }{R} \Bigg\{
1 - \frac{h}{R}  -\frac{\sigma_{cs} - \sigma_{as}}{\sigma_{ca}} 
\Bigg\}\le 0\;.
\end{eqnarray}
Noting that $cos \theta = 1 - h/R$, we observe that for a general variation $\delta h$
and $R>0$, the entropy maximum will be obtained based on the condition

\begin{align}
    \cos \theta -\frac{\sigma_{ls} - \sigma_{vs}}{\sigma_{lv}}  = 0\;,
\end{align}
which is Young's equation for the contact angle. The classical result is thereby obtained when the shape of the contact line is radially symmetric. This outcome provides evidence to demonstrate that the proposed topological concept is a general explicit geometric statement and also has a direct link to classical thermodynamics. 

\section{Computation of Macroscopic Contact Angle}
\begin{algorithm}
    \caption{Implementation in computation of macroscopic contact angles for droplets in multiphase system.}
    \label{codel}
    \begin{algorithmic}
    \For{each droplet in the system}
        \State 2D surface manifold generation by marching cubes algorithm
        \State Gaussian smooth for the manifold:
        \State $v_i' = v_i +\alpha \sum w_{ij}(v_j - v_i)$
        \For{each triangle on the manifold}
            \State Compute triangle area:
            \State $A_i = \frac{1}{2}|v_{12} \times v_{13}|$
            \State Compute principal curvatures $\kappa_1$ and $\kappa_2$
            \State Compute Gaussian curvature:
            \State $\kappa_{Gi} = \kappa_1 \kappa_2$
            \State Compute mean curvature:
            \State $\kappa_{Mi} = \frac{\kappa_1 + \kappa_2}{2}$
            \If {$\kappa_M < 0$}
                \State $\kappa_{Gi} A_i = - \kappa_{Gi} A_i$
            \EndIf
        \EndFor
        \State Compute $\theta_{macro}$ by normalizing deficit curvature with the number of contact line loop $N$:
        \State $\theta_{macro} = \lambda \Theta = \frac{4 \pi - \sum \kappa_{Gi} A_i}{4N}$
    \EndFor
    \end{algorithmic}
\end{algorithm}
\begin{figure}
\hspace*{-0.5cm}\includegraphics[width=0.5\textwidth]{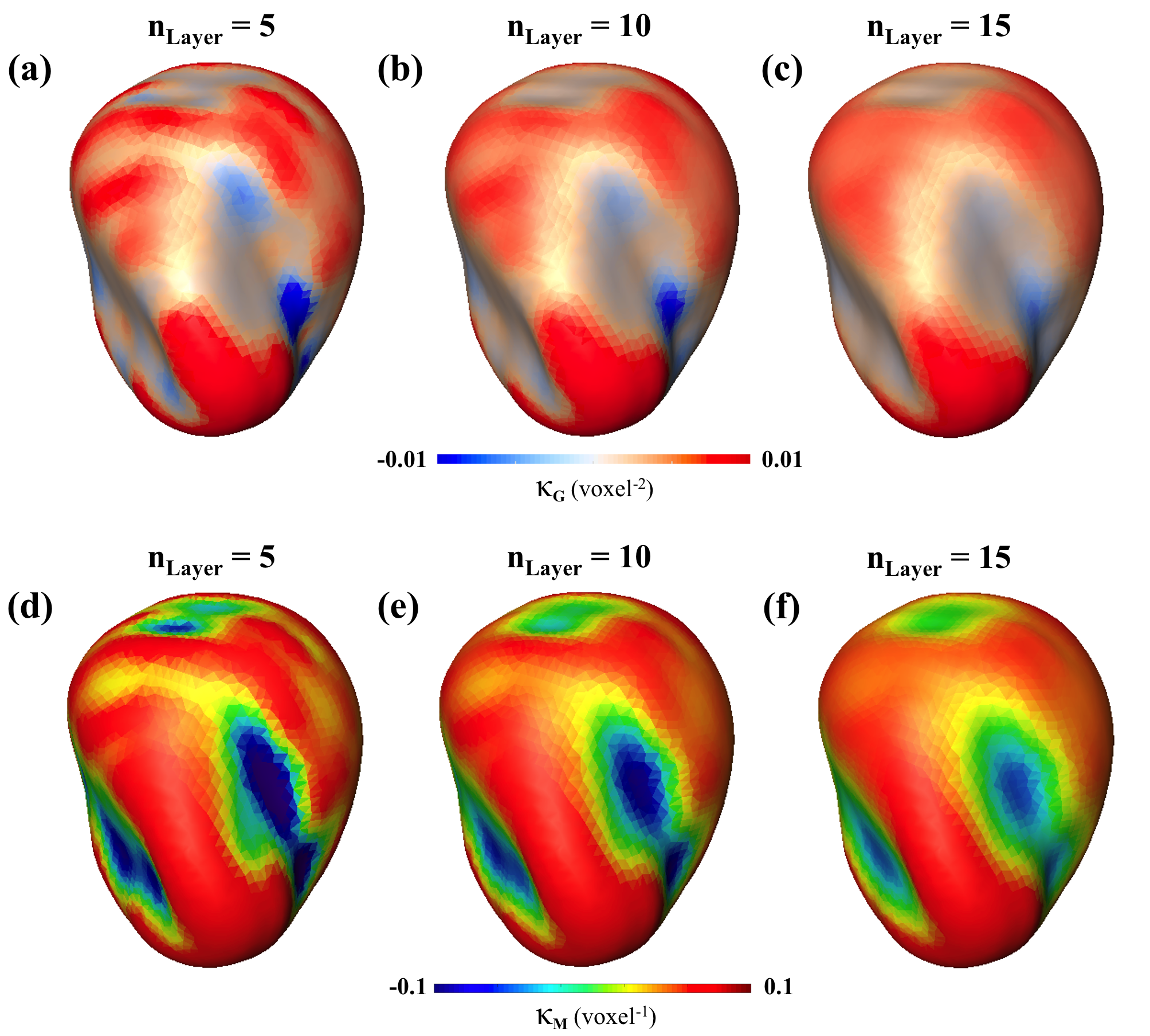}
\caption{The distribution of (a) - (c) Gaussian curvatures and (d) - (f) mean curvatures on the droplet interface for various degrees of $n_{Layer}$, which determines the number of triangles considered to be neighbours of a given point. Consequently, it determines the quadratic form equation of the surface patch. As the value of $n_{Layer}$ increases, the number of the surrounding triangles to be considered as neighbours increases, which leads to more accurate results.}
\label{fig3}
\end{figure}

To explain the implementation of the proposed topological principle to characterize the wetting state of the system, we first chose a droplet in a complex and confined domain from the segmented image for illustration. By modeling the whole droplet surface manifold using a generalized marching cubes algorithm \cite{hege1997generalized}, we performed a triangular approximation of the interface for each region (e.g., liquid/vapor and liquid/solid) to preserve the droplet topology. To obtain accurate results and remove extreme outliers of curvatures, a simplification of the triangles by applying an edge collapse algorithm for the droplet surface was required to omit small non-smooth regions. For instance, the small concave spots in the convex regions will be degenerated during the triangle contraction process. The area of each triangle $A_i$ can be computed by the cross product of the two adjacent vectors ($v_{12}$, $v_{13}$) formed by the vertices ($v_1$, $v_2$ and $v_3$), which is $A_i = \frac{1}{2}|v_{12} \times v_{13}|$. Further, we smoothed the surface manifold by shifting its vertices to minimize the voxelization effects and segmentation errors. Each new vertex position ($v_i'$) was shifted towards the average position of its neighbours ($v_j$) by the weights ($w_{ij}$). Therefore, the vector average for the vertex can be obtained,

\begin{equation}
    \Delta v_i=\sum w_{ij}(v_j - v_i),
\end{equation}
and the updated vertex position will be

\begin{equation}
    v_i' = v_i +\alpha \Delta v_i,
\end{equation}
where $\alpha$ represents a scale factor ranging from $0$ to $1$. We then computed the magnitude and direction of the principal curvatures ($\kappa_1$ and $\kappa_2$) by fitting a quadratic form equation on the desired surface patch and obtaining its corresponding eigenvalues and eigenvectors \cite{armstrong2012linking}. The Gaussian curvature ($\kappa_{Gi}$) and mean curvature ($\kappa_{Mi}$) for each triangle can thereby be obtained in terms of the principal curvatures,

\begin{equation}
\begin{split}
    \kappa_{Gi} &= \kappa_1 \kappa_2,\\
    \kappa_{Mi} &= \frac{\kappa_1 + \kappa_2}{2}.
\end{split}
\end{equation}
\begin{figure}
\hspace*{-0.5cm}\includegraphics[width=0.53\textwidth]{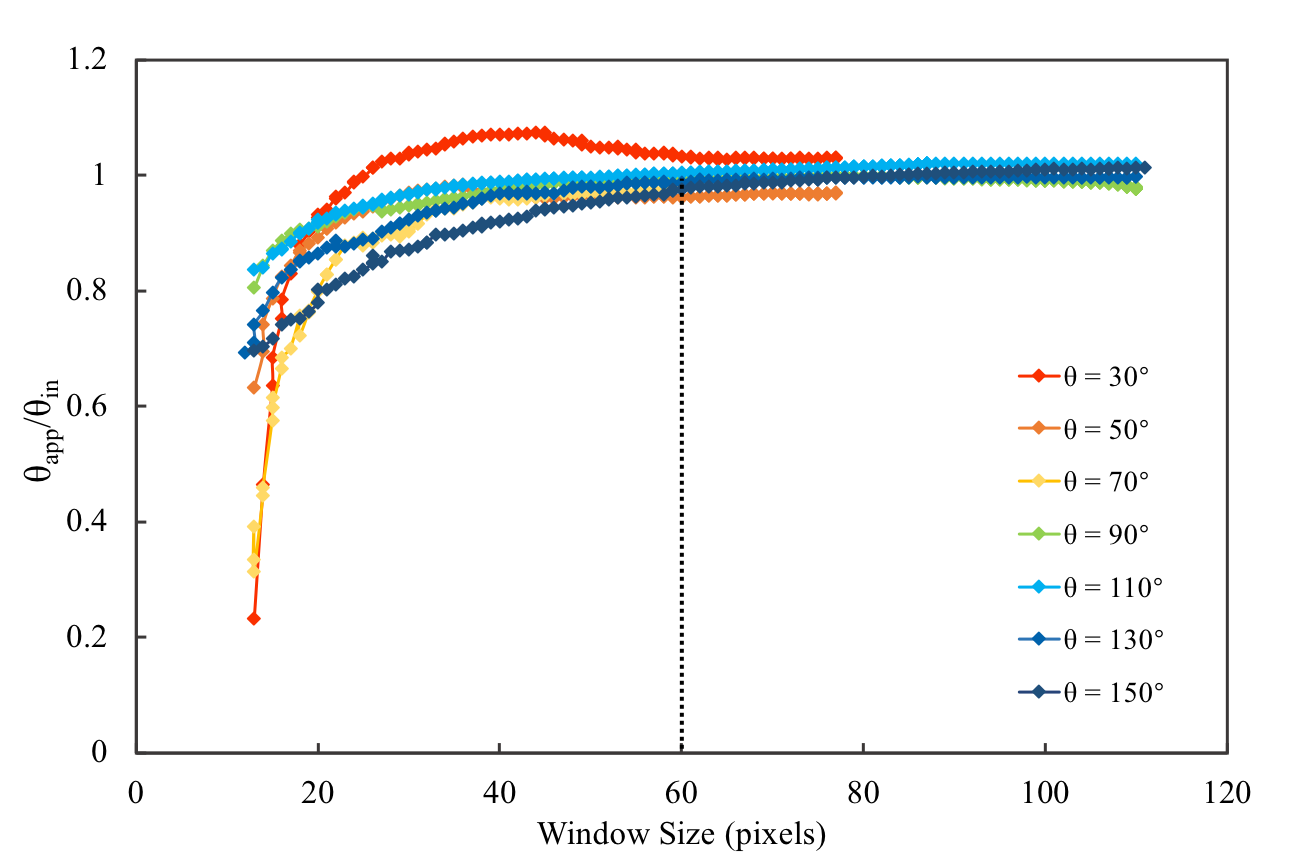}
\caption{Measured contact angles for different simulated droplets with increasing window size. It shows that an optimal window size of $60$ pixels is sufficient to accurately measure the apparent contact angle by the 2D local method.}
\label{fig4}
\end{figure}
The values of mean curvature provide the information of droplet interface, where the positive values of mean curvature indicate the convex interface and negative values indicate the concave interface as shown in Fig. \ref{fig3}d-f. In addition, there is a negative sign assigned to $\kappa_{Gi}A_i$ for concave interfaces and a value of $0$ assigned for flat interfaces. Finally, we can obtain the macroscopic contact angle, $\theta_{macro}$, for the droplet by normalizing the deficit curvature $\Theta$ as illustrated in Eq. (\ref{eq:4}),

\begin{equation}
    \theta_{macro} = \lambda \Theta = \frac{4 \pi - \sum \kappa_{Gi} A_i}{4N}
\end{equation}
where $N$ is the count of contact line loops. For more information, please see \cite{sun2020probing}. For multiphase systems where a large number of droplets are present, such as porous media and subsurface rocks, the proposed implementation procedure as described in Algorithm \ref{codel} enables the computation of macroscopic contact angles on a droplet-by-droplet basis.

\section{Results and Discussion}
\subsection{Wetting in Flat and Ideally-smooth Surfaces}

To unravel the effective and robust nature of the developed theory, we first compare the macroscopic contact angle $\theta_{macro}$ and the apparent microscopic contact angles $\theta_{app}$ on a flat and smooth surface. We performed a quasi-steady-state simulation of 3D sessile oil droplets immersed in immiscible ambient water on various flat surfaces that have different intrinsic contact angles defined by Young's equation. It is a more ideal solution for performing simulation than experiments where the droplet after deposition achieves a different contact angle, i.e. wetting hysteresis, when compared to the intrinsic contact angle. This is due to the difficulties in attaining experimental equilibrium condition and collecting ideally-flat surface \cite{arjmandi2017kinetics}. The equilibrium topologies of the oil droplets on the surface were simulated by minimizing the overall system energy and were subjected to constraints such as surface tension and gravitational energy. In the system, the overall energy is the sum of its interfacial potential energy and is governed by the Young-Laplace equation. For the interfacial potential energy of the droplet ($E$), it can be expressed as the sum of the respective interfacial energies,

\begin{equation}
    E=\int\int_{A_{ls}}\sigma_{ls}dA+\int\int_{A_{lv}}\sigma_{lv}dA+\int\int_{A_{vs}}\sigma_{vs}dA,
\label{eq:21}
\end{equation}
where $\sigma$ and $A$ represent interfacial tension and interfacial area, respectively. By applying Young's equation, Eq. (\ref{eq:21}) becomes

\begin{equation}
    \frac{E}{\sigma_{lv}}=A_{lv}-\int\int_{A_{ls}}\cos \theta_Y dA.
\end{equation}

\begin{figure*}
\centering\includegraphics[width=0.8\textwidth]{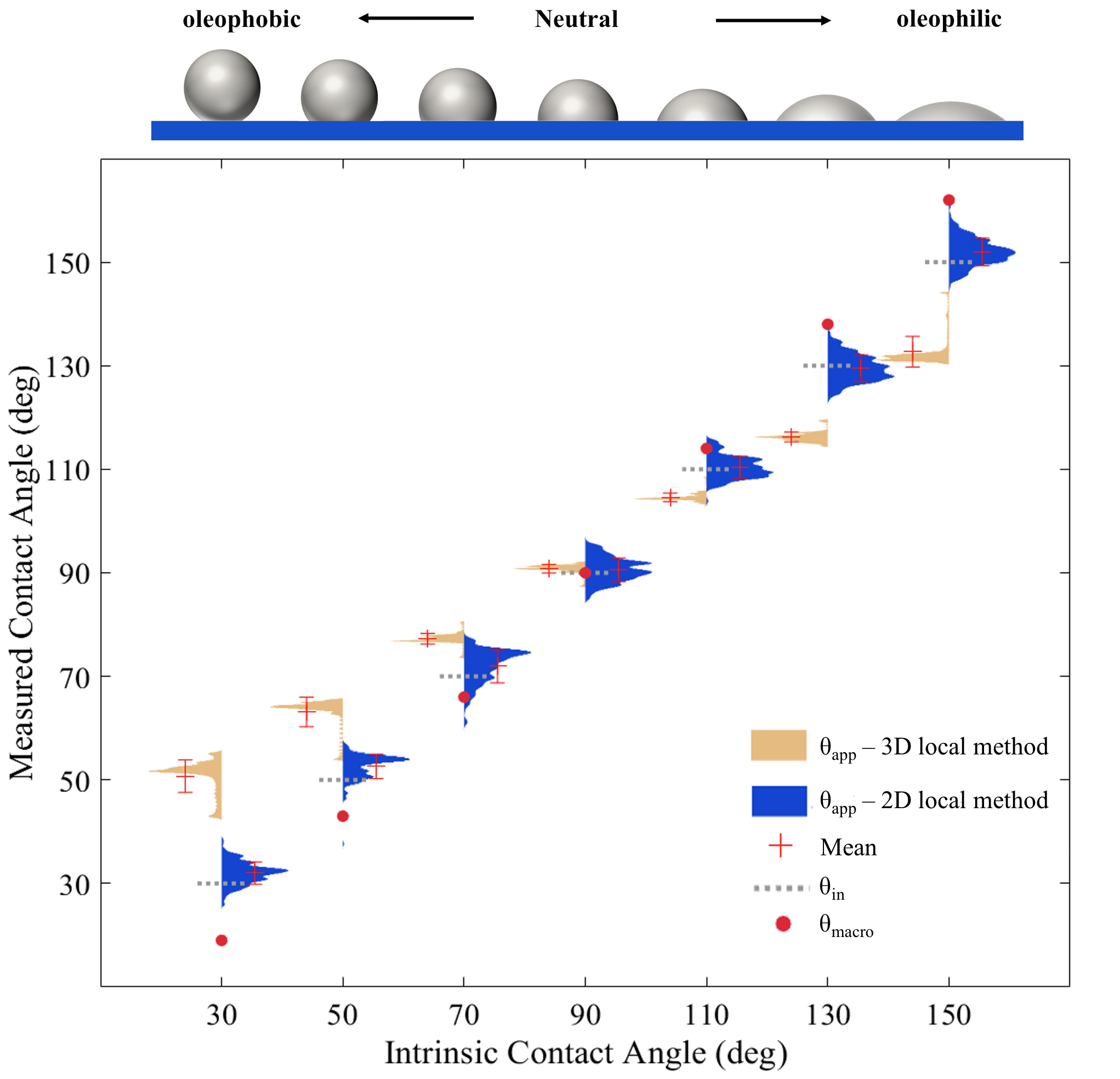}
\caption{The comparison diagram of $\theta_{macro}$ and the contact angle distributions of the 3D local measurement by \cite{alratrout2017automatic} and modified 2D local measurement by \cite{scanziani2017automatic} for the simulated droplets with different intrinsic contact angles.}
\label{fig5}
\end{figure*}

As shown in Fig. \ref{fig5}, the equilibrium topologies of oil droplets were simulated on a surface that gradually change from oleophobic to oleophilic. To obtain the microscopic apparent contact angles along the contact line, we implemented the 2D local method similar to \cite{scanziani2017automatic} and the 3D local method developed by \cite{alratrout2017automatic}. For the 2D local method, the contact points were smoothed using a moving average to determine the position and direction of the contact line. A 2D image was then extracted normal to the direction of the contact line for each contact point. In Fig. \ref{fig4}, we computed the contact angle measurement with different window (slice) sizes for each simulated droplet. The results demonstrate that a window (slice) size of $60$ pixels is sufficient to provide an accurate contact angle. In contrast to \cite{scanziani2017automatic}, both circular and linear regressions were applied to best fit the solid surface. The best fit of either the circular or linear regression was determined by the lower of the two root-mean-square deviations. If the circle approximation was selected as the best fit, a line tangent to the circle at the contact point was calculated to represent the slope of the surface at that point. In addition, a constant curvature of the liquid/vapor interface was assumed under the assumption that the system is at equilibrium. Consequently, the apparent contact angle was measured as the angle between the solid surface and liquid/vapor interface tangent lines \cite{meisenheimer2020optimizing}. For the 3D local method, Gaussian smoothing was applied to the droplet surface. Then, the two vectors that have a direction perpendicular to liquid/vapor and liquid/solid interface were identified for each contact point along the contact line. The apparent contact angle was thereby computed from the dot product of these vectors.  

\begin{figure*}
\centering\includegraphics[width=0.9\textwidth]{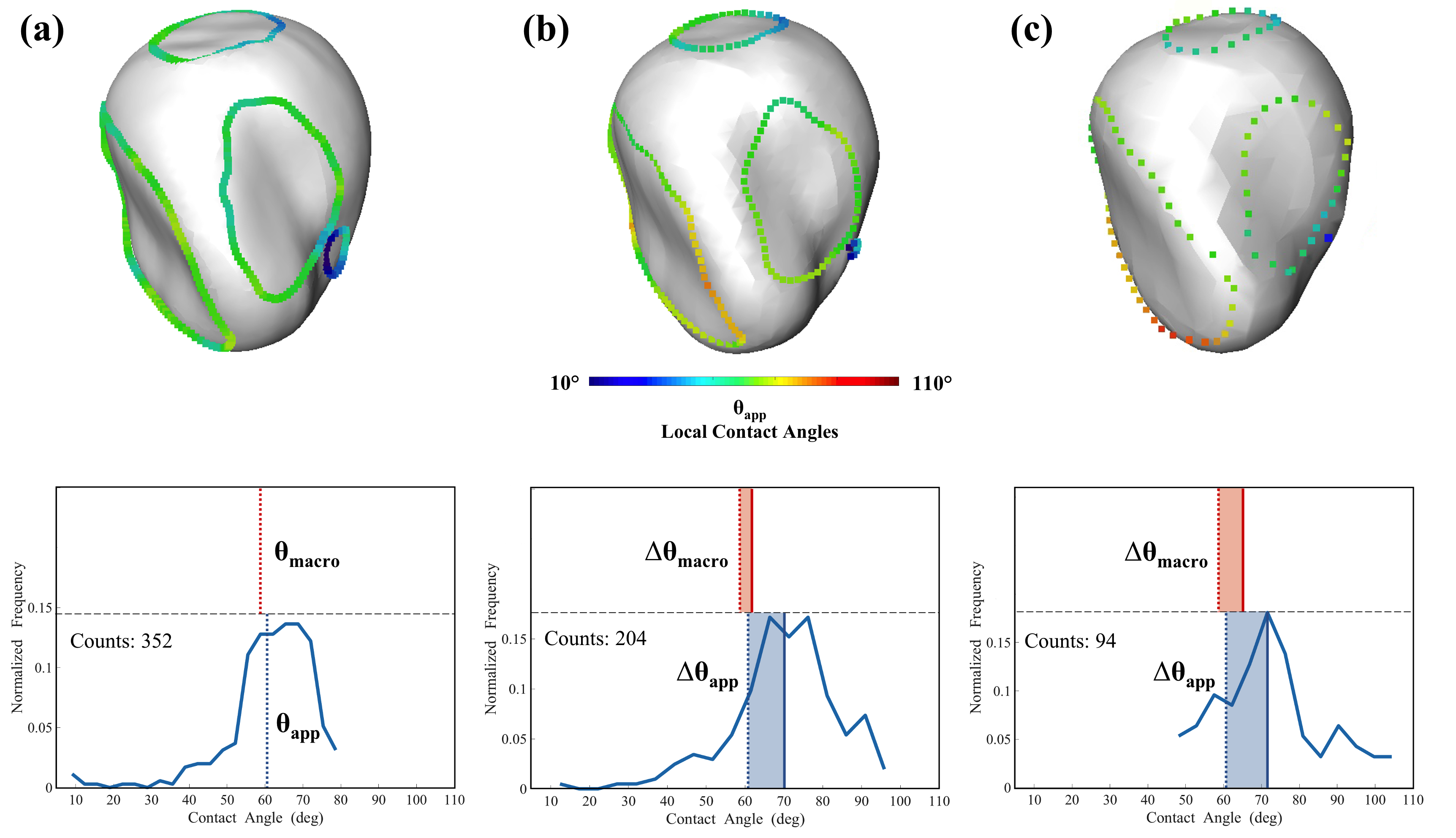}
\caption{Top: The comparison of droplet topology as down-sampling the image resolution with measured local contact angles for each three-phase contact points. Bottom: The corresponding macroscopic contact angle ($\theta_{macro}$) and 3D local measurements ($\theta_{app}$) associated with their change by comparing with the original resolution. As deficit curvature is less sensitive to resolution effects, it is particularly attractive as a way to capture wetting on rough surfaces where sub-resolution heterogeneity can have a demonstrated influence on the measured macroscopic contact angle.}
\label{fig6}
\end{figure*}

In Fig. \ref{fig5}, we compare the macroscopic contact angle $\theta_{macro}$ distributions from the developed topological principle to the apparent contact angle distributions by the two local methods for each intrinsic contact angle. It is shown that the 2D local measurement has a higher standard deviation compared with the 3D local measurement. However, the mean values of the contact angle distribution for the 2D local measurement are closer to the intrinsic contact angle $\theta_{in}$. Furthermore, the contact angle distribution calculated by 3D local method tends to have higher contact angles for more oleophobic surfaces and lower contact angles for more oleophilic surfaces, which provides a systematic bias (error) towards intermediate-wet conditions. Conversely, $\theta_{macro}$ is slightly less than the intrinsic contact angle for oleophobic surfaces and slightly greater than the intrinsic contact angle for oleophilic surfaces. The trend observed for $\theta_{macro}$ has a theoretical explanation since the total deficit curvature accounts for both local contact angles and the deformation of the contact line along the solid surface (line tension). The theoretical development for this particular aspect is explained in detail by Sun et al. \cite{sun2020probing}. Overall, contact angles measured for flat and homogeneous surfaces are comparable for all three tested methods, which supports the theoretical developments in Section 2.2 since the proposed topological approach under these ideal conditions can be simplified to Young's equation. 

\subsection{Wetting in Complex Geometries}
\begin{figure}
\centering\includegraphics[width=0.5\textwidth]{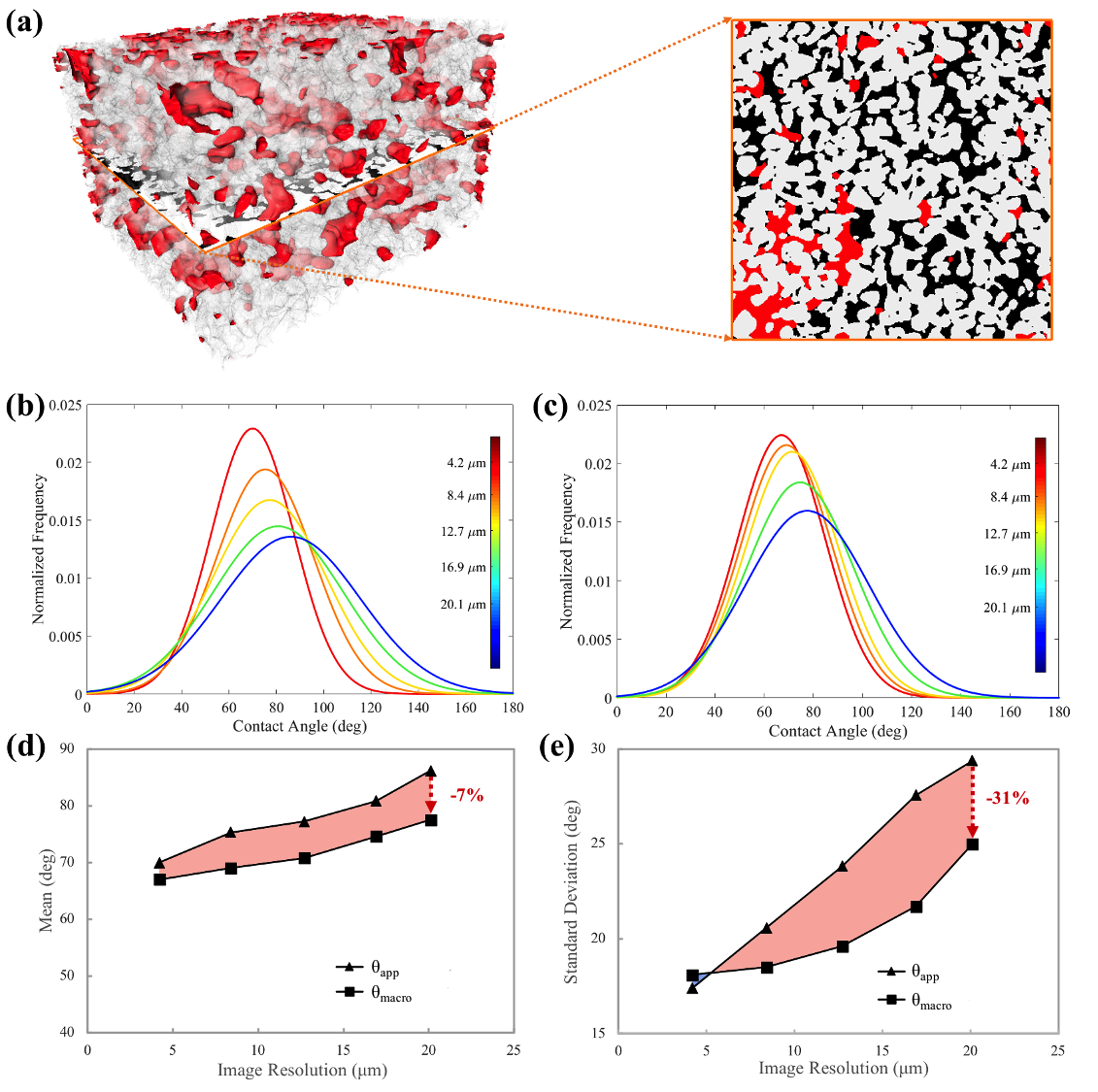}
\caption{The contact angle distributions of different image resolution for sintered glass by applying (b) 3D local method for each contact point and (c) the topological approach on a droplet-by-droplet basis. The plot of (d) mean and (e) standard deviation of the distribution for each image resolution.}
\label{fig7}
\end{figure}

\begin{figure}
\centering\includegraphics[width=0.5\textwidth]{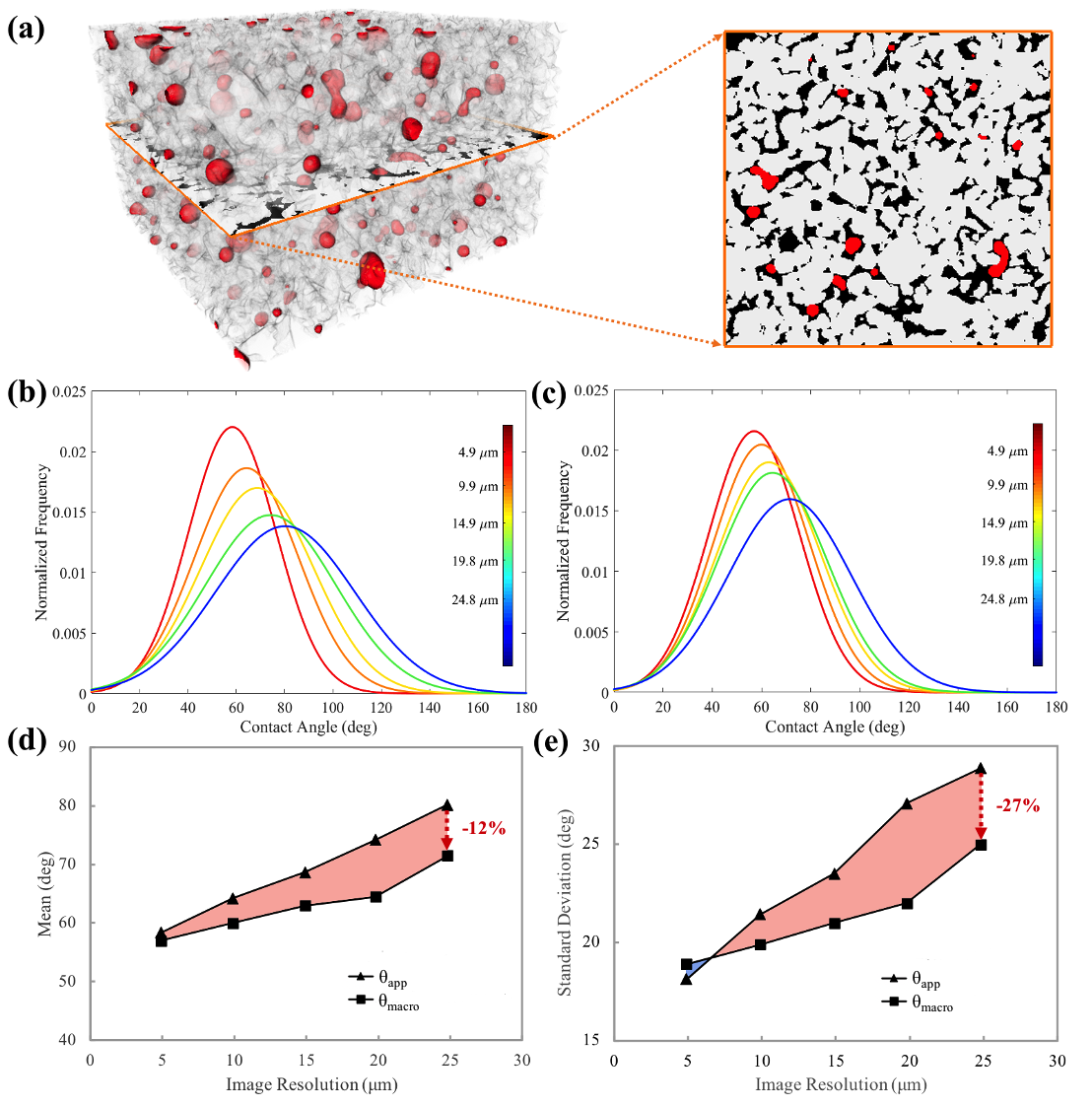}
\caption{The contact angle distributions of different image resolution for Bentheimer sandstone by applying (b) 3D local method for each contact point and (c) the topological approach on a droplet-by-droplet basis. The plot of (d) mean and (e) standard deviation of the distribution for each image resolution.}
\label{fig8}
\end{figure}

To further investigate the wetting state on heterogeneous surfaces where surface chemistry and roughness are present, we performed two quasi-static capillary forces dominated flow experiments for sintered glass at the Swiss Light Source and Bentheimer sandstone at the Australian National University \cite{dataset}. The details of the experimental setup and image processing can be found in \cite{schluter2016pore} and \cite{sun2020probing}. The sintered glass has a relatively lower degree of surface geometry with smooth surfaces and a lower degree of surface chemistry. While for Bentheimer sandstone, the surface chemical heterogeneity, surface geometry and roughness are relatively higher. In Table \ref{tab}, a summary of experimental setup and image processing is given for both sintered glass and Bentheimer sandstone. The 3D fluid configurations and spatial arrangements for both samples are displayed in Figs. \ref{fig7}a and \ref{fig8}a.

\begin{table*}
   \centering
   \caption{Summary of experimental setup and image processing.}
   \begin{tabular}{L{2.7cm}C{5cm}C{5cm}} 
      \toprule 
        & Sintered Glass & Bentheimer Sandstone\\
      \midrule
      Porosity & $31.8 \%$ & $24.1 \%$\\
      Permeability & $21.5 \pm 2$ $\rm D$ & $4.3$ $\rm D$\\
      Sample diameter & $4$ $\rm mm$ & $4.9$ $\rm mm$\\
      Sample length & $10$ $\rm mm$ & $10$ $\rm mm$\\
      Non-wetting phase & n-decane & ambient air\\
      Wetting phase & brine & brine\\
      Injection rate & $0.1$ $\rm \mu L/min$ & $0.3$ $\rm \mu L/min$\\
      Brine saturation & $78 \%$ & $93 \%$\\
      Imaging source & fast synchrotron-based tomography & bench-top helical micro-tomography\\
      Image resolution & $4.2$ $\rm \mu m$ & $4.95$ $\rm \mu m$\\
      Time step & $30$ $\rm s$ & $1.35$ $\rm hr$\\
    \bottomrule
    \end{tabular}
   \label{tab}
\end{table*}

To investigate the resolution effects, we first performed a resolution study by picking a well-resolved droplet from the dataset of Bentheimer sandstone. The original dimension of image for the droplet is $47 \times 51 \times 48$ voxels. We then down-sample the image of the droplet by halving the resolution twice as shown in Fig. \ref{fig6}a-c. It is intuitive in Fig. \ref{fig6} how the down-sampling impacts the droplet interfacial curvature and contact line loops. The results demonstrate that the macroscopic contact angle based on the deficit curvature is less sensitive to resolution effects than local contact angle measurements, since the change in the mean value of $\theta_{app}$ is larger than the change in $\theta_{macro}$. Local contact angle measurements cannot be performed without resolving the contact line region. It is clear that the resolution necessary to measure the angles along the contact line is lost before the topological structure of the contact line loops is destroyed. In Fig. \ref{fig6}a-c, we investigate that the local contact angles vary significantly as the resolution decreases. For the topological approach based on the Gauss-Bonnet theorem, there remains a basis to obtain deficit curvature when the topological structure of the contact line loops is captured, even if only a single voxel represents the contact line loop as shown on the bottom right of Fig. \ref{fig6}c.

Further, we performed the contact angle distribution measurements for both samples on a collection of droplets by applying the topological approach for each droplet and 3D local method for each contact point. By gradually decreasing the image resolution as shown in Figs. \ref{fig7} and \ref{fig8}, the changes in the mean and standard deviation of the contact angle distributions for the topological approach are notably less than that of the local measurements. For sintered glass, the mean of the contact angle distribution decreases by an additional $7 \%$ while there is an additional $31 \%$ reduction in the standard deviation for the topological approach. Similar results for the topological method are investigated in the Bentheimer sandstone with a decrease in mean contact angle and standard deviation of $12 \%$ and $27 \%$, respectively. Overall, at sufficiently high resolution, the topological approach and local measurements are comparable. While at lower resolutions, the topological approach deviates less from the high resolution results than the local contact angle measurements. 

\begin{figure}
\hspace*{-0.5cm}\includegraphics[width=0.5\textwidth]{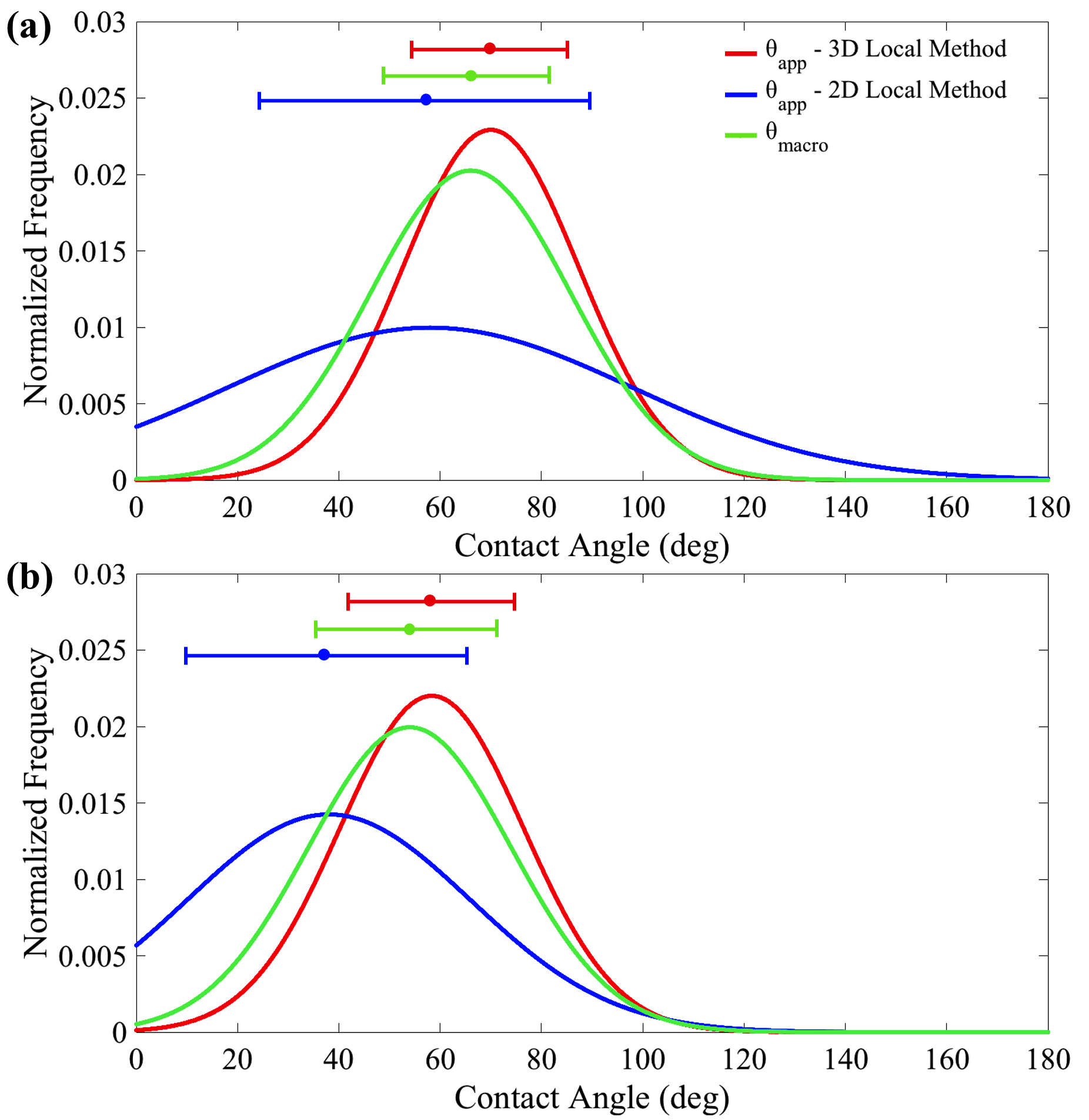}
\caption{The contact angle distributions for (a) sintered glass and (b) Bentheimer sandstone by applying topological approach, 2D and 3D local methods with the highest resolution obtained together with the associated mean and standard deviation.}
\label{fig9}
\end{figure}

Lastly, we performed contact angle measurements for both samples by applying the topological approach, 2D and 3D local methods with the highest image resolution obtained as shown in Fig. \ref{fig9}. It reveals that the 2D local measurement provides a lower mean and a higher standard deviation of the contact angle distribution. However, the mean value of the 3D local measurement is the highest, which can be explained in the simulation of droplets results for the flat and homogeneous surfaces. For more oleophobic (water-wet) surfaces, contact angle measurements computed by the 3D local method is larger than the intrinsic contact angles. The $\theta_{macro}$ distribution computed by the topological approach falls between the two local methods. Overall, the topological and 3D local methods are comparable at sufficiently high resolution, suggesting that microscopic wetting information is captured. 

\section{Conclusions}

Recent methods for characterizing wetting behavior rely on restrictive thermodynamic laws such as Young's equation for flat homogeneous surfaces, and Wenzel and Cassie-Baxter models for heterogeneous surfaces \cite{young1805iii,wenzel1936resistance,cassie1944wettability}. In this work, we developed a new theory based on topological principles to determine the wetting state of multiphase systems. Specifically, the concept of thermodynamic equilibrium is not necessary since the formulation is a geometrical statement between the total surface curvature and global topology of a fluid object. In this regard, the proposed methodology does not require visualization of the three-phase contact line. It thus would be advantageous for applications where a large fluid body is resolved while the contact line pinning occurs at a smaller length scale. This could be particularly informative for problems where the droplet is connected to a thin film, such as a droplet slipping along a declining plane and/or the corner flow mechanism observed in porous media flows. We also highlight that the proposed theory is consistent with Young's equation for flat and homogeneous surfaces by performing a variational analysis.

By comparing with traditional local contact angle measurements \cite{alratrout2017automatic,scanziani2017automatic,klise2016automated}, we assess the sensitivity and robustness of the proposed topological approach. It is found that the proposed theory has less sensitivity to resolution effects than the tested local methods. While at sufficiently high image resolution, the results are comparable. As observed in Fig. \ref{fig6}, the resolution necessary to measure the contact line is lost before the topological structure of the contact line loops are destroyed. Contact angles cannot be measured without resolving the contact line region. However, the contact line region must always form a loop even if it is captured by only one voxel. Therefore, there remains a basis to measure $\theta_{macro}$ even at low image resolution since topology and interfacial curvatures of fluid surface are still captured. Another way to think about why  $\theta_{macro}$ works well at low resolution is because fluid blobs are physically larger than the contact lines, and we can use information embedded in the fluid interface structure to infer the wetting state.

Overall, the results demonstrate that the proposed theory provides an accurate macroscopic wetting description based on microscopic wetting information that is less susceptible to resolution-based errors due to its robust nature of measuring interfacial area and curvature rather than angles along the contact line. The theoretical links of the proposed theory explored in other recent publications \cite{sun2020linking,sun2020probing} are further highlighted by the variational analysis presented herein. The particular example was provided for a simple sessile droplet system. However, more complex systems with morphologically and mineralogically heterogeneous geometries could be considered leading to a more general condition for the constrained entropy inequality. This could be particularly important for the development of multiphase flow models that explicitly account for wettability and will be the focus of further work. 

\section*{Acknowledgments}
A part of this work was performed at the Swiss Light Source, Paul Scherrer Institute, Villigen, Switzerland. C. S. acknowledges an Australian Government Research Training Program Scholarship. A. H. acknowledges ARC DE180100082 and the ANU/UNSW Digicore Research Consortium. J. M. acknowledges an award of computer time provided by the Department of Energy Early Science Program. This work was supported by the U.S. National Science Foundation, Hydrologic Sciences Program under award No.1344877. This research also used resources of the Oak Ridge Leadership Computing Facility, which is a DOE Office of Science User Facility supported under Contract DE-AC05-00OR22725.




\end{document}